\documentclass[acmsmall,nonacm]{acmart}
\setcopyright{none}
\renewcommand\footnotetextcopyrightpermission[1]{}
\usepackage[utf8]{inputenc}
\usepackage{natbib}
\usepackage{amsmath}
\usepackage{amsfonts}
\usepackage{mathpartir}
\usepackage{newunicodechar}
\usepackage{xspace}
\usepackage[frozencache=true,cachedir=minted-cache]{minted} 
\newunicodechar{ℕ}{\ensuremath{\mathbb{N}}}
\newunicodechar{₁}{\ensuremath{_\texttt{1}}}
\newunicodechar{∀}{\ensuremath{\forall}}
\newunicodechar{λ}{\ensuremath{\lambda}}
\newunicodechar{∈}{\ensuremath{\in}}
\newunicodechar{∣}{\ensuremath{\mid}}
\newunicodechar{≡}{\ensuremath{\equiv}}
\newunicodechar{≤}{\ensuremath{\le}}
\newunicodechar{′}{\ensuremath{'}}
\newunicodechar{⊎}{\ensuremath{\uplus}}

\newcommand{\eg}{e.g.,\xspace}

\newcommand{\denote}[1]{\ensuremath{\llbracket{#1}\rrbracket}}

\newcommand{\ag}[1]{\mintinline{agda}{#1}}

\begin{document}

\title{Toward SMT-based Refinement Types in Agda}

\author{Gan Shen}
\affiliation{\institution{University of California, Santa Cruz} \country{USA}}
\author{Lindsey Kuper}
\affiliation{\institution{University of California, Santa Cruz} \country{USA}}

\begin{abstract}
  Dependent types offer great versatility and power, but developing proofs with them can be tedious and requires considerable human guidance.  We propose to integrate Satisfiability Modulo Theories (SMT)-based refinement types into the dependently-typed language Agda in an effort to ease some of the burden of programming with dependent types and combine the strengths of the two approaches to mechanized theorem proving.
\end{abstract}

\maketitle

\section{Introduction}
Agda~\citep{norell-agda} is a dependently-typed programming language where types can contain (depend on) terms.
With the expressivity of dependent types, we can encode rich, precise properties of programs using types and prove that those properties hold at compile time through type checking.
A standard example of a dependent type is the type of lists of a given length: \ag{Vec A n}.
Here \ag{A} is the type of list elements and \ag{n} is the length of the list.
Then we can write a \ag{lookup} function of type \ag{Vec A n → Fin n → A} where \ag{Fin n} is the type of natural numbers less than \ag{n}, and rest assured that there will never be an out-of-bounds list index error at run time.

However, the power of \emph{full} dependent types comes at a price --- writing programs amounts to writing proofs, and proofs can be difficult and tedious to write.
For instance, suppose we want to use the aforementioned \ag{Fin n} in a context where \ag{Fin (n + 1)} is required.
Even though something of type \ag{Fin n} should be usable in that context, the Agda type checker complains, forcing us to write a proof to show that a term of type \ag{Fin n} can also be typed \ag{Fin (n + 1)}.
This problem could be avoided if we had Satisfiability Modulo Theories (SMT)-based refinement types~\citep{rondon-liquid-types, liquid-haskell, refinement-reflection, f-star}, which have a subtyping rule that uses an off-the-shelf SMT solver to automate the proof for us.
We can think of refinement types as a restricted form of dependent types in which types can contain terms drawn from a decidable logic.
As a result, proofs can be done automatically by SMT solvers, making certain programs easier to write.

In this paper, we propose to integrate SMT-based refinement types into the dependently-typed language Agda in an effort to ease the burden of programming with dependent types. We illustrate some of the challenges of using dependent types, discuss how refinement types can alleviate some of them, and consider implementation approaches.

\section{The Burden of Proof}

To motivate the need for refinement types in Agda, we consider a simple example from the Agda standard library~\citep{agda-fin} (with some tweaks to aid readability).
The type \ag{Fin n} denoting natural numbers less than \ag{n} is inductively defined as:\footnote{In Agda, arguments wrapped in curly brackets are implicit arguments, which means we can omit them when calling the function because the type checker can infer their values.}

\begin{minted}{agda}
data Fin : ℕ → Set where
  fzero : {n : ℕ} → Fin (suc n)
  fsuc  : {n : ℕ} (i : Fin n) → Fin (suc n)
\end{minted}

\noindent Here \ag{Fin n} is a data type indexed by a natural number\footnote{\ag{ℕ} is the type of natural numbers in Agda.  Following Peano arithmetic, it has two constructors, \ag{zero} and \ag{suc}.  Agda also supports decimal notation, so one can write \ag{2} for \ag{suc (suc zero)}.} \ag{n}.  Elements of \ag{Fin n} can be seen as natural numbers in the set $\{m \ | \ m < n \}$.

We can understand this data type definition as saying:

\begin{itemize}
\item For all natural numbers \ag{n}, \ag{fzero} is a member of the set \ag{Fin (suc n)}.
\item For all natural numbers \ag{n} and \ag{i} in set \ag{Fin n}, \ag{fsuc i} is a member of set \ag{Fin (suc n)}.
\end{itemize}

Even though semantically \ag{Fin n} is just a specialized \ag{ℕ}, because it is a whole separate definition, one cannot use it in a context where a \ag{ℕ} is required.
This could potentially lead to unnecessary verbose proofs and poorly performing code.  For example, consider the predecessor function \ag{pred} for \ag{Fin n}, also from the Agda standard library, which assumes the predecessor of \ag{fzero} is still \ag{fzero}:

\begin{minted}{agda}
pred : {n : ℕ} → Fin n → Fin n
pred fzero    = fzero
pred (fsuc i) = inject₁ i
\end{minted}

Everything looks pretty normal except that we have to call this mysterious \ag{inject₁} function.
To understand what is happening, let's take a close look at how the \ag{fsuc i} case is type checked.  Due to the type of \ag{fsuc}, we know that \ag{fsuc i} must have type \ag{Fin (suc n′)} for some \ag{n′}, so we know that \ag{n = suc n′} and our end goal changes to \ag{Fin (suc n′)}.

To get the predecessor of \ag{fsuc i}, we want to simply drop the \ag{fsuc} and return \ag{i}, but \ag{i} has type \ag{Fin n′} as opposed to \ag{Fin (suc n′)}, as required.
Naturally, a natural number less than \ag{n′} is also less than \ag{suc n′}, but Agda stubbornly rejects that.  Instead, we need to provide a proof saying a \ag{Fin m} is also a \ag{Fin (suc m)}, which is called \ag{inject₁} in the Agda standard library:

\begin{minted}{agda}
inject₁ : {m : ℕ} → Fin m → Fin (suc m)
inject₁ fzero    = fzero
inject₁ (fsuc i) = suc (inject₁ i)
\end{minted}
\ag{inject₁} takes a \ag{Fin m} and lifts it to \ag{Fin (suc m)} while keeping the underlying data structure intact.
From a programming perspective, \ag{inject₁} does nothing interesting, but has time complexity $O(n)$.
It is acceptable if we never intend to actually execute the code, but would be vexing for those who want to extract a efficient implementation from Agda or who just want to do practical programming in Agda.

\section{Refinement Types to the Rescue}

Having seen how cumbersome it is to use the inductively defined \ag{Fin n} type in Agda, we turn to a different encoding that uses subtype polymorphism and an underlying SMT solver to ease the burden of programming.
A natural number less than \ag{n} is really just a subset of natural numbers.  With refinement types, we can say:
\[ \mathit{Fin} \ n \triangleq \{ x:\mathbb{N} \ | \ x < n \} \]
In general, base refinements of the form $\{ x:B \ | \ P(x) \}$ are the building blocks of refinement types.  Here $B$ is a basic type, and $P$ is the refinement predicate constraining the value $x$.
This refinement type denotes a subset of $B$ whose elements satisfy the refinement predicate $P$.
Thus, a basic type $B$ without refinement can be thought of as an abbreviation for $\{ x:B \ | \ \top \}$.

We often find ourselves wanting to use a more specific type in a more relaxed context --- as in the above \ag{pred} example, where we want to use \ag{Fin n′} in a context where \ag{Fin (suc n′)} is required --- so we define a subtyping rule for refinement types:
\[
\inferrule
    {\denote{\Gamma, x:B} \vDash \denote{P(x)} \Rightarrow \denote{Q(x)}}
    {\Gamma \vdash \{x:B \ | \ P(x) \} \preceq \{x:B \ | \ Q(x) \}} \quad \textsc{$\preceq$-Base}
\]
The \textsc{$\preceq$-Base} rule says that the base refinement $\{ x:B \ | \ P(x) \}$ is a subtype of $\{ x:B \ | \ Q(x) \}$ under the context $\Gamma$ if and only if $\denote{P(x)}$ implies $\denote{Q(x)}$ under the context $\denote{\Gamma}$ extended with $\denote{x : B}$.
Here $\denote{\cdot}$ means the translation from Agda terms to SMT terms.

For example, the subtyping relation:
\[ n : \mathit{ℕ} \vdash \{x:\mathit{ℕ} \ | \ x < n\} \preceq \{x:\mathit{ℕ} \ | \ x < (suc \ n)\} \]
follows from the validity of the implication:
\[ n : \mathit{ℕ},\, x : \mathit{ℕ} \vDash x < n \Rightarrow x < n + 1 \]
which tells us $\mathit{Fin \ n}$ is a subtype of $\mathit{Fin \ (suc \ n)}$.
This is what we need to be able to infer a sufficiently specific return type for \ag{pred}.

\section{Implementing Refinement Types in Agda}
A refinement type $\{ x:B \ | \ P(x) \}$ can be readily emulated in a dependent type system as the $\Sigma$ (dependent pair) type $\Sigma_{x : B}\,P(x)$, where $B$ is a type and $P$ is a function from $B$ to a type.
For readability, following the Agda standard library~\citep{agda-refinement}, we define a record type \ag{Refinement} in Agda and give it a convenient syntax resembling set comprehension:
\begin{minted}{agda}
record Refinement (B : Set) (P : B → Set) : Set where
  constructor _,_
  field
    value  : B
    .proof : P value

syntax Refinement B (λ x → P) = [ x ∈ B ∣ P ]
\end{minted}
A value of type \ag{Refinement} is a pair of \ag{value} and \ag{proof}, where \ag{value} is the basic type being refined and \ag{proof}\footnote{The dot syntax marks the \ag{proof} component of \ag{Refinement} as \emph{proof-irrelevant}, which prevents us from using the proof in computation; we discuss this further below.} is the refinement.
To construct a value of refinement type, we use the \ag{_,_}\footnote{Agda supports defining mixfix operators using underscores \ag{_} to mark where arguments are positioned when applied, so we can apply \ag{_,_} to \ag{a} and \ag{b} by writing \ag{a , b}.} constructor and provide both value and proof:
\begin{minted}{agda}
2≡2  : [ x ∈ ℕ ∣ x ≡ 2 ]
2≡2  = 2 , refl

2≤10 : [ x ∈ ℕ ∣ x ≤ 10 ]
2≤10 = 2 , s≤s (s≤s z≤n)
\end{minted}
A function that operates on values of refinement type can extract the value and proof from the argument and use them to build the result:
\begin{minted}{agda}
pred : {n : ℕ} → [ x ∈ ℕ ∣ x < n ] → [ x ∈ ℕ ∣ x < n ]
pred (zero  , p) = zero , p
pred (suc x , p) = x    , <-trans (n<1+n x) p
\end{minted}

The refinement type syntax is convenient, but the real advantage of SMT-based refinement type systems like Liquid Haskell~\citep{liquid-haskell} and F*~\citep{f-star} is that they automate away much of the burden of having to explicitly construct the proof.
Instead of explicitly writing proofs like \ag{refl}, \ag{s≤s (s≤s z≤n)}, and \ag{<-trans (n<1+n x) p}, we can ask an SMT solver to carry out the proof for us.

Furthermore, if such a proof does not exist --- that is, if the SMT formula translated from the Agda refinement predicate turns out not to be valid --- the solver can provide a counterexample, corresponding to a model that satisfies the negation of the formula.  Such a counterexample could provide more insight into the problem than a plain type mismatch error message.
  
To bridge between Agda and SMT, we plan to use Schmitty~\citep{schmitty}, an Agda library that provides bindings for SMT-LIB~\citep{barrett-smtlib} (the common input language used by all mainstream SMT solvers) compatible solvers and macros for translating Agda terms to SMT-LIB scripts.  Combined together, these features can enable automatic proving.
In particular, Schmitty provides a macro \ag{solveZ3} that calls an external SMT solver during type checking.  \ag{solveZ3} works by:
\begin{enumerate}
\item Getting the typing context $\Gamma$ and the expected type $\tau$ from its call site.
\item Translating the implication $\Gamma \implies \tau$ to SMT-LIB script and asking if the negation of the formula is unsatisfiable (which is equivalent to the validity of the original formula in classical logic; we will discuss the interplay between this use of the solver and Agda's constructive logic below).
\item Inserting a dummy proof (a \emph{postulate} in Agda) at the call site if the solver deems the negated formula to be unsatisfiable, and otherwise reporting an error (counterexample).
\end{enumerate}
With the help of the \ag{solveZ3} macro, the above code examples can be simplified as follows:
\begin{minted}{agda}
2≡2  : [ x ∈ ℕ ∣ x ≡ 2 ]
2≡2  = 2 , solveZ3

2≤10 : [ x ∈ ℕ ∣ x ≤ 10 ]
2≤10 = 2 , solveZ3

pred : {n : ℕ} → [ x ∈ ℕ ∣ x < n ] → [ x ∈ ℕ ∣ x < n ]
pred (zero  , _) = zero , solveZ3
pred (suc x , _) = x    , solveZ3
\end{minted}

We can introduce further macros that hide the proof component of values of type \ag{Refinement}.
The resulting code feels similar in spirit to programming in a mainstream functional programming language like Haskell or OCaml, yet programmers still have access to the full power of dependent types if need be.
Tricky proofs that cannot be automated by the underlying SMT solver can still be carried out in Agda by hand.

There are disadvantages to using an SMT solver to automate proofs in Agda.  Two that are especially worthy of mention are:
\begin{itemize}
\item
  The dummy proof generated by \ag{solveZ3} doesn't have computational meaning. We mitigate this by annotating the proof component of \ag{Refinement} as \emph{proof-irrelevant}, which prevents us from using the proof in computation, so whether it has computational meaning doesn't matter.
  It's still an open problem to extract proofs from a solver and reify them into Agda \citep{barrett2015proofs}.
\item
  SMT solvers reason in classical logic, so by using them to automate proofs in Agda, we are essentially strengthening Agda's logic from constructive to classical.
  That said, it is already possible to constructively prove an excluded-middle property (without the use of SMT) for many of the predicates we are working with (\eg for equality of natural numbers, we can prove that for all natural numbers \ag{x} and \ag{y}, \ag{x ≡ y ⊎ ¬(x ≡ y)}), so it can be argued that in such cases SMT only introduces automation rather than changing the underlying logic.
\end{itemize}

\section{Conclusion and Future Work}

Theorem-proving systems have historically fallen into two camps~\citep{boutin-reflection}: interactive theorem proving,  exemplified by proof assistants such as Agda, and automated theorem proving, exemplified by SMT solvers and tools that build on them.
Each approach has its strengths and weaknesses.
SMT-based tools provide ``push-button'' operation: given a proposition, they return either true or false without human guidance.
They are great as far as they go, and can automate many tedious proofs,
but they offer little feedback to the user.
On the other hand, dependent-types-based tools provide great versatility and power, but
developing proofs with them requires considerable human guidance.

The boundary between the two camps is blurred by hybrid automatic/interactive (sometimes called ``auto-active''~\citep{leino-auto-active}) verification approaches, with different tools providing varying degrees of granularity of feedback provided to the human user, and different styles of specifying program properties.
A version of Agda augmented with SMT-based refinement types would represent one point in this  space, as do emerging proof assistants that integrate SMT automation such as Lean~\citep{de-moura-lean} and F*~\citep{f-star}.  Unfortunately, the research community's knowledge of such hybrid verification tools is fragmented: they are dots on a map, but we have no robust theory to tie them together, leading to the tools being poorly understood and hence under-exploited.
As a starting point, we propose that integrating SMT-based refinement types into Agda would result in a hybrid verification tool in which the strengths of SMT-based and dependent-types-based tools can be combined.
In the long run, we hope to not explore not only this one point in the design space, but to systematically survey the landscape of such hybrid tools and contribute to a holistic scientific understanding of the space.

\bibliography{references}

\end{document}